\documentclass[11pt,a4paper]{article}

\usepackage[margin=1in]{geometry}
\usepackage{amsmath,amssymb,amsthm}
\usepackage[utf8]{inputenc}
\usepackage[T1]{fontenc}
\usepackage{graphicx}
\usepackage{cite}
\usepackage{IEEEtrantools}
\usepackage{hyperref}

\newcommand{\ack}[1]{%
  \section*{Acknowledgements}%
  #1%
}

\makeatletter
\newcommand{\titlehead}[1]{}

\newcommand{\@address}{}
\newcommand{\@subject}{}
\newcommand{\@keywords}{}
\newcommand{\@corres}{}

\newcommand{\address}[1]{\gdef\@address{#1}}
\newcommand{\subject}[1]{\gdef\@subject{#1}}
\newcommand{\keywords}[1]{\gdef\@keywords{#1}}
\newcommand{\corres}[1]{\gdef\@corres{#1}}
\newcommand{\email}[1]{\href{mailto:#1}{#1}}

\renewcommand{\@maketitle}{%
  \newpage
  \null
  \vskip 2em%
  \begin{center}%
    {\LARGE \bfseries \@title \par}%
    \vskip 1.5em%
    {\large
      \lineskip .5em%
      \begin{tabular}[t]{c}%
        \@author
      \end{tabular}\par}%
    \vskip 1em%
    {\small \@address \par}%
  \end{center}%
  \vskip 1.5em%
}

\newcommand{\printmetadata}{%
  \vspace{1em}\noindent
  \ifx\@subject\@empty\else\textbf{Subject Areas:} \@subject \\ \fi
  \ifx\@keywords\@empty\else\textbf{Keywords:} \@keywords \\ \fi
  \ifx\@corres\@empty\else\textbf{Contact:} \@corres \fi
}
\makeatother

\begin{document}

\bstctlcite{IEEEexample:BSTcontrol}

%%%% Article title
\title{SIESTA-QCOMP: Bridging Classical and Quantum Computing for Strongly
Correlated Systems}

%%%% Author details
\author{Y. Pouillon$^{1}$, F. Marchesin$^{1}$, Ivano Tavernelli$^{2}$,
E. Artacho$^{1,3,4,5}$}

%%%% Author address
\address{$^{1}$CIC nanoGUNE, Tolosa Hiribidea 76, 20018 Donostia-San Sebasti\'{a}n, Spain\\[2pt]
$^{2}$IBM Research Z\"{u}rich, S\"{a}umerstrasse 4, CH-8803 R\"{u}schlikon,
Switzerland\\[2pt]
$^{3}$Donostia International Physics Center (DIPC), Paseo Manuel de
Lardizabal 4, 20018 San Sebasti\'{a}n, Spain\\[2pt]
$^{4}$Ikerbasque, Basque Foundation for Science, 48011 Bilbao, Spain\\[2pt]
$^{5}$Theory of Condensed Matter, Cavendish Laboratory, University of
Cambridge, J. J. Thomson Ave, Cambridge CB3 0US, United Kingdom}

%%%% Subject entries
\subject{computational physics, quantum computing, atomic and molecular
physics, chemical physics}

%%%% Keyword entries

%%%% Corresponding author
\corres{Yann Pouillon (\email{ygm.pouillon@nanogune.eu})}

\maketitle

%%%% Abstract text to be placed here %%%%%%%%%%%%
\begin{abstract}
Density Functional Theory (DFT) is the established workhorse of computational
materials science, yet it faces fundamental challenges with strongly
correlated electrons, such as in transition metal complexes and bond-breaking
processes. While quantum computing offers a theoretically exact path to
resolve these problems, current hardware limitations prohibit fully quantum
simulations of realistic material systems. To bridge this gap, we present
SIESTA-QCOMP, a modular suite embedding quantum methodologies within DFT.
Harnessing SIESTA's localised basis set, we implement a projection-based
embedding scheme that treats the bulk environment classically while mapping
the strongly correlated "impurity" subspace to a quantum processor. The
architecture uses a socket-based driver to orchestrate workflows between the
Fortran-based SIESTA engine and Python-based frameworks like Qiskit, ensuring
high interoperability. We detail the hybrid workflow, currently utilising a
Variational Quantum Eigensolver (VQE), and discuss the road map toward methods
such as Sample-based Quantum Diagonalisation (SQD). We illustrate the
framework on an hypothetical iron porphyrin calculation and argue that such
hybrid architectures are not merely interim solutions but essential
foundations for integrating quantum utility into materials design, as
envisioned by our broader QCOMP4DFT initiative.

\printmetadata
\end{abstract}
%%%%%%%%%%%%%%%%%%%%%%%%%%%

\vspace{1.5em}

{\it
\noindent
Submitted to: Philosophical Transactions of the Royal Society A \\
Status: Accepted, Under Press \\
Manuscript date: 2026-06-30 \\
}

%%%%%%%%%% Insert the texts which can accommodate on first page in the tag "fmtext" %%%%%

%\begin{fmtext}
%\end{fmtext}

%%%%%%%%%%%%%%% End of first page %%%%%%%%%%%%%%%%%%%%%

\maketitle

\newpage

\section{Introduction}
\label{sec:introduction}

Density Functional Theory (DFT) offers a great balance between computational
cost and accuracy, which has allowed researchers to predict the properties of
thousands of compounds, revolutionising fields from catalysis to
photovoltaics. However, as the community pushes the boundaries of materials
design, the limitations of standard DFT approximations have become
increasingly apparent \cite{Teale2022}. The "standard model" of electronic
structure fails significantly when applied to strongly correlated systems ---
materials where electron-electron interactions are too complex to be averaged
out by a mean-field potential. This failure is most acute in systems
containing localised $d$- and $f$-electrons, such as transition metal oxides,
high-temperature superconductors, and metal-organic compounds fundamental to
life sciences \cite{Hait2021}. \\

In principle, Full Configuration Interaction (FCI) methods can solve these
correlation problems, but they scale exponentially with system size, rendering
them intractable for all but the smallest molecules on classical computers.
This is the domain where quantum computing holds its most significant promise.
As originally envisioned by Feynman, quantum computers can simulate quantum
systems efficiently, offering a path to exact solutions for electronic
structure problems that are classically impossible \cite{Feynman1982}.
In particular, the dimension of the Hilbert space that can be explored grows
exponentially with the number of qubits. However, contemporary quantum
hardware is limited by low qubit counts, short coherence times, and
significant gate noise \cite{Preskill2018}. Consequently, the dream of
simulating a complex biological molecule, such as haemoglobin, entirely on a
quantum processor remains a distant goal. This creates a dichotomy: we have
traditional methods (DFT) that can handle large systems but fail at
correlations, and quantum methods that handle correlations but fail at system
size. \\

To overcome these respective limitations, the field is converging towards
hybrid quantum-classical architectures. One of the most promising strategies
is quantum embedding. Conceptually, this approach partitions a material into
two regions: a large environment, which is treated efficiently by DFT, and a
smaller, strongly correlated active space (or "impurity"), which is solved on
a quantum processor \cite{Rossmannek2023,Battaglia2024}. In most cases, the
quantum algorithm itself is hybrid as well, to further mitigate current
hardware limitations and make a more efficient use of surrounding HPC
resources \cite{Alexeev2024}. This "best of both worlds" approach relies on
the assumption that strong correlation is often a local phenomenon. By
defining a high-level quantum solver for the active space and coupling it
self-consistently with a mean-field approach, we can extend the reach of
quantum computers to realistic sizes of materials without requiring millions
of qubits. Over the years, different classes of quantum-centric algorithms
suitable for material and molecular sciences, such as the Variational Quantum
Eigensolver (VQE) \cite{Peruzzo2014,McClean2016}, the Sample-based Krylov
Quantum Diagonalisation (SKQD) \cite{Robledo2025}, and SqDRIFT
\cite{Piccinelli2026}, have emerged. \\

In this paper, we introduce SIESTA-QCOMP, a novel implementation of
projection-based embedding \cite{Lee2019} that integrates the widely used
SIESTA DFT code \cite{Soler2002,Garcia2020} with quantum computing back-ends.
SIESTA is uniquely suited for this task due to its use of Numerical Atomic
Orbitals (NAOs). Unlike plane-wave codes, the basis functions of SIESTA are
naturally localised in real space, simplifying the definition and projection
of local active spaces required for embedding. SIESTA-QCOMP is distinguished
by its modular architecture. Rather than embedding quantum routines directly
into the Fortran codebase of SIESTA, we utilise a socket-based communication
driver. This decouples the classical engine from the rapidly evolving quantum
software ecosystem --- primarily Python-based frameworks like Qiskit ---
ensuring that improvements in quantum algorithms can be integrated without
altering the source code on the classical side. \\

The remainder of this paper is structured as follows.

\begin{itemize}
  \item Section \ref{sec:best-of-both-worlds} discusses the theoretical
  underpinnings of hybrid quantum-classical workflows, analysing the strengths
  and bottlenecks of current algorithms like the Variational Quantum
  Eigensolver (VQE).
  \item Section \ref{sec:scheme} details the specific implementation of the
  projection embedding scheme within SIESTA, defining the mathematical
  framework for the subspace projection.
  \item Section \ref{sec:architecture} describes the SIESTA-QCOMP software
  architecture, its socket-based interface and the workflow to address systems
  such as the iron porphyrin complex.
  \item Section \ref{sec:quantum-utility} outlines the long-term vision of the
  QCOMP4DFT initiative. We argue that achieving "quantum utility" requires
  not just hardware improvements, but a long-lasting community-wide effort to
  build interoperable, robust software ecosystems that seamlessly blend the
  classical and quantum domains.
\end{itemize}

\bigskip

\section{Hybrid Quantum-Classical Approaches: The "Best of Both Worlds"?}
\label{sec:best-of-both-worlds}

To understand the necessity of hybrid architectures, one must first confront
the limits of the classical workhorse. Density Functional Theory (DFT) has
arguably been the most successful method in the history of computational
physics, offering an unparalleled balance between computational cost and
accuracy \cite{Singh2021,Kumar2024,Tang2025}. However, its reliance on Mean-Field Theory and a
single-configuration reference creates fundamental blind spots. While DFT
excels at describing ground-state structural properties, it suffers from
intrinsic errors --- notably self-interaction error and delocalisation error
--- which become critical in regimes of strong electron correlation
\cite{Verma2020,Zois2024}. In
practice, these theoretical deficiencies manifest as severe qualitative
failures in specific material classes:

\begin{itemize}
\item Strongly Correlated Systems: Systems with localised d- or f-electrons,
such as transition metal oxides, are often described as metallic when they are
experimentally insulating \cite{Maniar2025}.
\item Organic Metallochemistry: The description of 3d metals within organic
scaffolds (e.g., organometallic catalysts) is frequently inaccurate, leading
to incorrect spin-state ordering \cite{Saiz2020}.
\item Bond Breaking: The dissociation of chemical bonds, such as the C-N bond
in butyronitrile, involves multi-reference character that standard functionals
cannot capture \cite{Rossmannek2023}.
\item Excited States: Time-Dependent DFT (TDDFT) often fails to correctly order
excited-state energies, particularly for charge-transfer excitations
\cite{Baranova2025}.
\end{itemize}

These failures all stem from the non-interacting picture inherent to the
Kohn-Sham ansatz. To correct this, one must reintroduce the explicit many-body
wavefunction, but doing so for the entire system is in general computationally
prohibitive. \\

The solution lies in quantum embedding: treating the bulk of the system at the
DFT level while treating the problematic "impurity" (the defect, the metal
centre, or the breaking bond) with high-level quantum mechanics. When starting
our endeavour, we considered four primary embedding strategies:

\begin{enumerate}
  \item Projection-based Embedding: The system is partitioned based on the
  electronic density. Orthogonality between the subsystems is enforced via
  localisation procedures or level-shift projectors \cite{Rossmannek2023}.
  \item Range-Separated Embedding: The partition is based on interactions,
  separating short-range dynamic correlations from long-range static
  correlations \cite{Rossmannek2021}.
  \item Effective Hamiltonian Embedding: The partition is energetic. High-energy
  states are "downfolded" into renormalised parameters, in a conceptually
  similar way to QM/MM approaches \cite{Bauman2019}.
  \item Density-Matrix Embedding Theory (DMET): The system is partitioned into
  spatial fragments coupled to an effective cluster through a minimal set of
  bath orbitals \cite{Kirsopp2022}.
\end{enumerate}

For the SIESTA-QCOMP project, we have prioritised Projection-based Embedding
(PE) \cite{Lee2019}. This choice is driven both by technical
suitability and strategic collaborations. First, the use of localised
Numerical Atomic Orbitals (NAOs) in SIESTA makes density projection natural
and numerically stable. Second, we aim to complement rather than duplicate
ongoing community efforts; with parallel collaborations already focusing on
range-separation \cite{Rossmannek2021} and DMET \cite{Shajan2025}, projection
embedding offers a distinct and interoperable pathway. It allows us to
rigorously define an active space A embedded in an environment B without
requiring the complex bath optimisation of DMET \cite{Ralli2024}. \\

Once the problem is mapped to the active space, we can employ the Variational
Quantum Eigensolver (VQE) to resolve the active space. While VQE is widely
used on near-term noisy quantum computing devices, it is not without
challenges. Recent benchmarking \cite{Carreras2025} and theoretical analysis
\cite{Vaquero-Sabater2025} highlight significant hurdles:

\begin{itemize}
\item Barren Plateaus: The optimisation landscape often becomes flat as system
size increases, making gradient-based navigation difficult.
\item Measurement Overhead: The non-commuting nature of the Hamiltonian terms
requires a vast number of measurements ("shots") to achieve chemical precision.
\item Ansatz Validity: Choosing a parametrised circuit that spans the relevant
chemical space without introducing excessive noise remains an open problem.
\end{itemize}

Despite these challenges, the projection-based VQE approach provides the
necessary test bed for developing the software infrastructure that will
eventually support fault-tolerant algorithms. We are currently reproducing
previous results \cite{Rossmannek2023} with SIESTA-QCOMP.

\section{Theoretical Framework: Projection-Based Embedding}
\label{sec:scheme}

The SIESTA-QCOMP workflow utilizes projection-based embedding \cite{Manby2012,
Lee2019,Ralli2024} to partition a large physical system into two distinct regions: a
localized "active" region ($A$), containing the strongly correlated electrons
(e.g., $d$-orbitals of a transition metal), and the "environment" ($B$),
comprising the remainder of the system. This approach allows the environment
to be treated at the mean-field DFT level while the active region is solved
using high-accuracy quantum algorithms. \\

A critical prerequisite for embedding is the spatial definition of the active
region. Unlike plane-wave methods that require complex localization
transformations (e.g., Wannier functions), SIESTA utilizes Numerical Atomic
Orbitals (NAOs) that are strictly localized in real space. This "natural"
locality allows for a direct partition of the Hilbert space based on atomic
indices or orbital shells, which forms the cornerstone of our approach. \\

The formal process begins with a converged Kohn-Sham (KS) DFT calculation of
the full system to obtain the total density matrix, $D_{tot} = D_A + D_B$.
Here, $D_A$ is the projection of the density onto the basis functions spanning
the active atoms, and $D_B$ is the remainder. To transition from the
non-orthogonal NAO basis to the second-quantized form required by quantum
algorithms, we construct an orthogonalized basis for the active subspace
(typically through a Löwdin or canonical transformation of the $A$-block). \\

The effective electronic Hamiltonian for the active space, $\hat{H}_{emb}$, is
expressed as:

\begin{equation}
\label{eqn:second-quantised-hamiltonian}
\hat{H}_{emb} = \sum_{p,q \in A} h_{pq}^{eff} \hat{a}^\dagger_p \hat{a}_q
+ \frac{1}{2} \sum_{p,q,r,s \in A} g_{pqrs} \hat{a}^\dagger_p \hat{a}^\dagger_q \hat{a}_r \hat{a}_s
\end{equation}

where $p, q, r, s$ denote the orthogonalized active-space orbitals. The core
of the embedding logic resides in the one-electron integrals, $h_{pq}^{eff}$:

\begin{equation}
\label{eqn:effective-hamiltonian}
h_{pq}^{eff} = h_{pq}^{core} + [V_{H}(D_B)]_{pq} + [V_{xc}(D_{tot})]_{pq} + [\mu \hat{P}_{B}]_{pq}
\end{equation}

In this expression, $V_{H}(D_B)$ represents the Hartree potential of the
environment, accounting for the classical electrostatic repulsion from region
$B$. The exchange-correlation potential, $V_{xc}(D_{tot})$, is evaluated using
the \textit{total} density to capture the non-linear coupling between regions,
ensuring the active electrons feel the correct chemical environment. To
enforce the Pauli exclusion principle and prevent the active electrons from
"collapsing" into states already occupied by the environment, we include a
Mu-shift projection operator, $\hat{P}_{B} = S D_B S$ (where $S$ is the
overlap matrix). By setting the parameter $\mu$ to a large value (e.g., $10^6$
Hartree), the environment states are pushed to high energies, effectively
removing them from the active space optimization. \\

Once $\hat{H}_{emb}$ is constructed, it is mapped to qubit operators
($\hat{H}_{qubit}$) via the Jordan-Wigner or Bravyi-Kitaev transformations
\cite{Jordan1928, Tranter2018}. This transforms the fermionic
creation/annihilation operators into Pauli strings:

\begin{equation}
\label{eqn:pauli-strings}
\hat{H}_{qubit} = \sum_i c_i P_i, \quad P_i \in \{I, X, Y, Z\}^{\otimes n}
\end{equation}

In SIESTA-QCOMP, the ground state of $\hat{H}_{qubit}$ is determined using the
ADAPT-VQE algorithm \cite{Grimsley2019}. This variational approach prepares a
trial wavefunction $|\psi(\vec{\theta})\rangle$ on the Quantum Processing Unit
(QPU) and iteratively minimizes the energy expectation value:

\begin{equation}
\label{eqn:energy-vqe}
E(\vec{\theta}) = \langle \psi(\vec{\theta}) | \hat{H}_{qubit} | \psi(\vec{\theta}) \rangle
\end{equation}

By offloading only the high-complexity correlation problem to the QPU while
retaining the chemical environment via SIESTA's efficient DFT implementation,
we enable the simulation of complex materials and molecular systems that were
previously computationally inaccessible.

\section{Architecture and Implementation: The SIESTA-QCOMP Workflow}
\label{sec:architecture}

A significant challenge in realising hybrid quantum-classical workflows is the
software engineering barrier. State-of-the-art electronic structure  codes
like SIESTA are often highly optimised, monolithic codebases, written in
compiled languages (e.g., Fortran), whereas the quantum computing ecosystem is
rapidly evolving and predominantly Python-based (e.g., Qiskit, Cirq,
Pennylane). Merging these two worlds by directly embedding quantum libraries
into the Fortran source code is neither scalable nor maintainable. To address
this, SIESTA-QCOMP adopts a loosely coupled, modular architecture, that we
describe in section~\ref{sec:quantum-utility}. The design philosophy is a
strict "separation of concerns": the DFT code handles the environment and
heavy-duty integral processing, while a high-level Python driver orchestrates
the quantum logic and workflow control between the classical and quantum
environments. \\

Most quantum algorithms have been heavily influenced by quantum chemistry and
require a Hartree-Fock-like starting guess. While this means going beyond DFT,
we were able to take benefit from a synergistic collaboration through the
implementation of hybrid exchange-correlation functionals \cite{Becke1993}
into SIESTA. Since the latter introduce Hartree-Fock exchange in the DFT
workflow and rely on the computation of Coulomb Electron Repulsion Integrals
(ERIs), we could build our implementation on top of it. However, the way ERIs
are computed requires the use of intermediate Gaussian basis sets and induces
a few challenges regarding the stability and convergence rate of the
Self-Consistent Field (SCF) cycle \cite{Pouillon2026}. \\

The backbone of this architecture is a socket-based communication interface.
Instead of linking libraries at compile time, SIESTA and the quantum driver
communicate via internet sockets (UNIX or INET) during runtime. In this
model, the execution flow is inverted compared to standard inputs:

\begin{enumerate}
  \item The driver (Python script): Acts as the master orchestrator. It
  initialises the workflow, defines the embedding parameters, and first
  listens for data from the classical engine. Once all the ingredients to
  build a quantum ansatz have been received, the driver runs the quantum
  module and sends back its results to the classical engine.
  \item The classical engine (SIESTA): Provides the heavy computational
  resources. It performs the standard DFT cycle and, upon reaching specific
  hook points, pauses after sending data to the driver through the socket.
  Once the results from the quantum module are available, it requests the
  corresponding data from the driver and resumes the calculation.
  \item The quantum module (language-agnostic): Provides a specific Application
  Programming Interface (API) to the driver. It includes the embedding method
  at it highest level, which executes a (hybrid) quantum algorithm forming an
  intermediate level, which then relies on a quantum runtime to execute the
  critical part of the quantum computation. Once done, it returns the results
  to the driver.
\end{enumerate}

The hybrid simulation proceeds through a multi-step pipeline, automating the
projection-based embedding scheme described in Section~\ref{sec:scheme}.
The process begins with a standard DFT calculation in SIESTA. The system
converges the electronic density matrix of the full system ($D_{total}$) using
a standard functional (e.g., PBE or LDA). This provides the mean-field
environment density. \\

\begin{figure}
  \centerline{\includegraphics[width=4.5in]{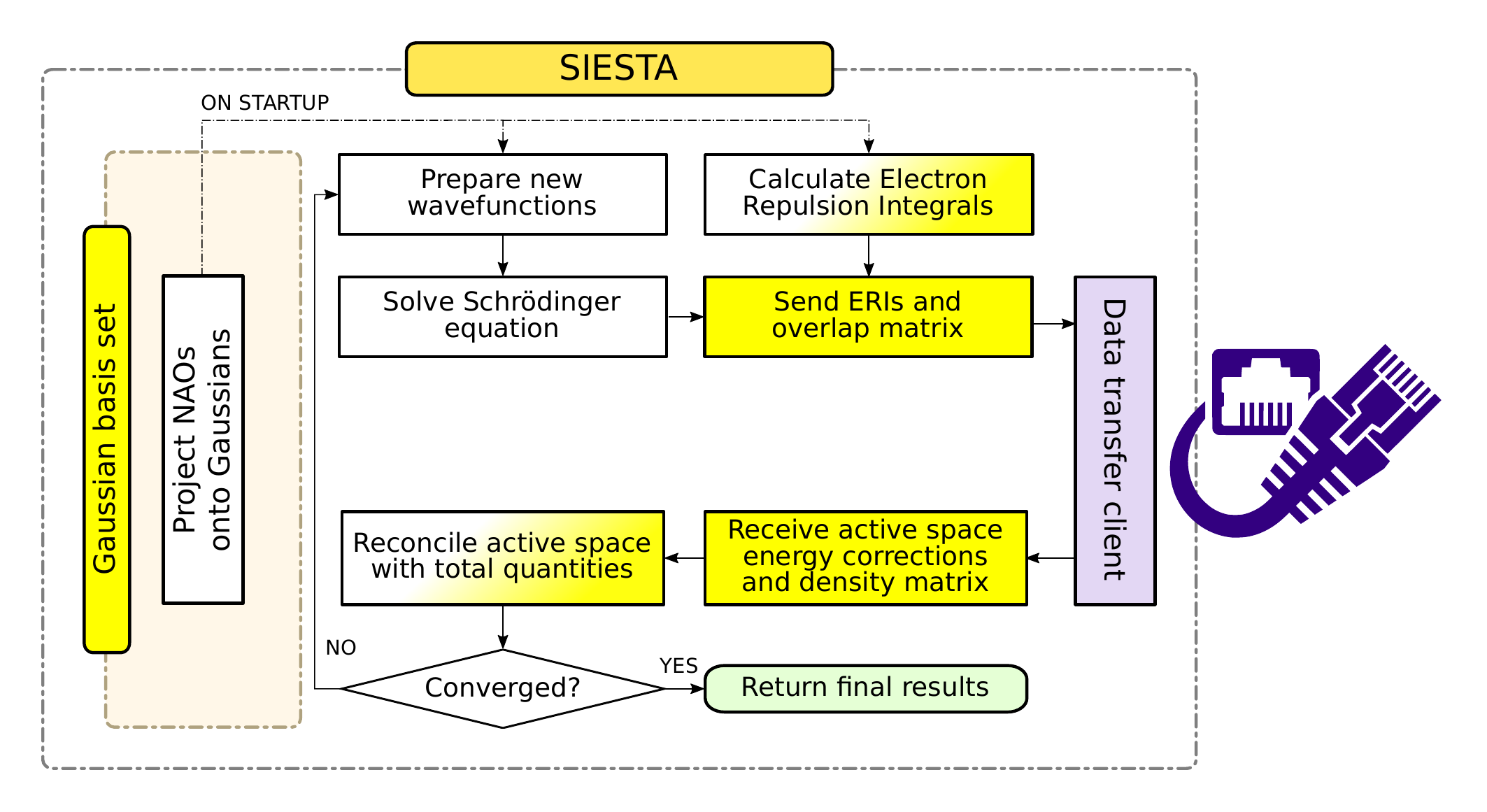}}
  \caption{Classical part of the hybrid workflow, managed by SIESTA. The
  Gaussian basis sets required to calculate ERIs are provided by the hybdrid
  exchange-correlation subsystem of SIESTA. The SCF cycle is then opened at
  specific hook points where the data transfer can happen. Cells with a white
  background correspond to the original SIESTA SCF implementation, while a
  yellow background shows the new components. Steps involving modified SIESTA
  components are represented with a white-yellow gradient.
  \label{fig:workflow-classical}}
\end{figure}

Based on the user-defined active space (e.g., "Iron 3d orbitals"), SIESTA
computes the ingredients for the quantum ansatz. Crucially, SIESTA relies on
its efficient sparse matrix operations to calculate the 1-electron integrals
($h_{pq}^{eff}$) and on advanced filtering mechanisms to calculate the minimum
possible number of 2-electron Coulomb integrals ($g_{pqrs}$). SIESTA
serialises these Hamiltonian parameters and transmits them via the socket to
the Python driver, before waiting for the quantumly computed results. This
part of the workflow is illustrated in Fig.~\ref{fig:workflow-classical}. \\

\begin{figure}
  \centerline{\includegraphics[width=4.5in]{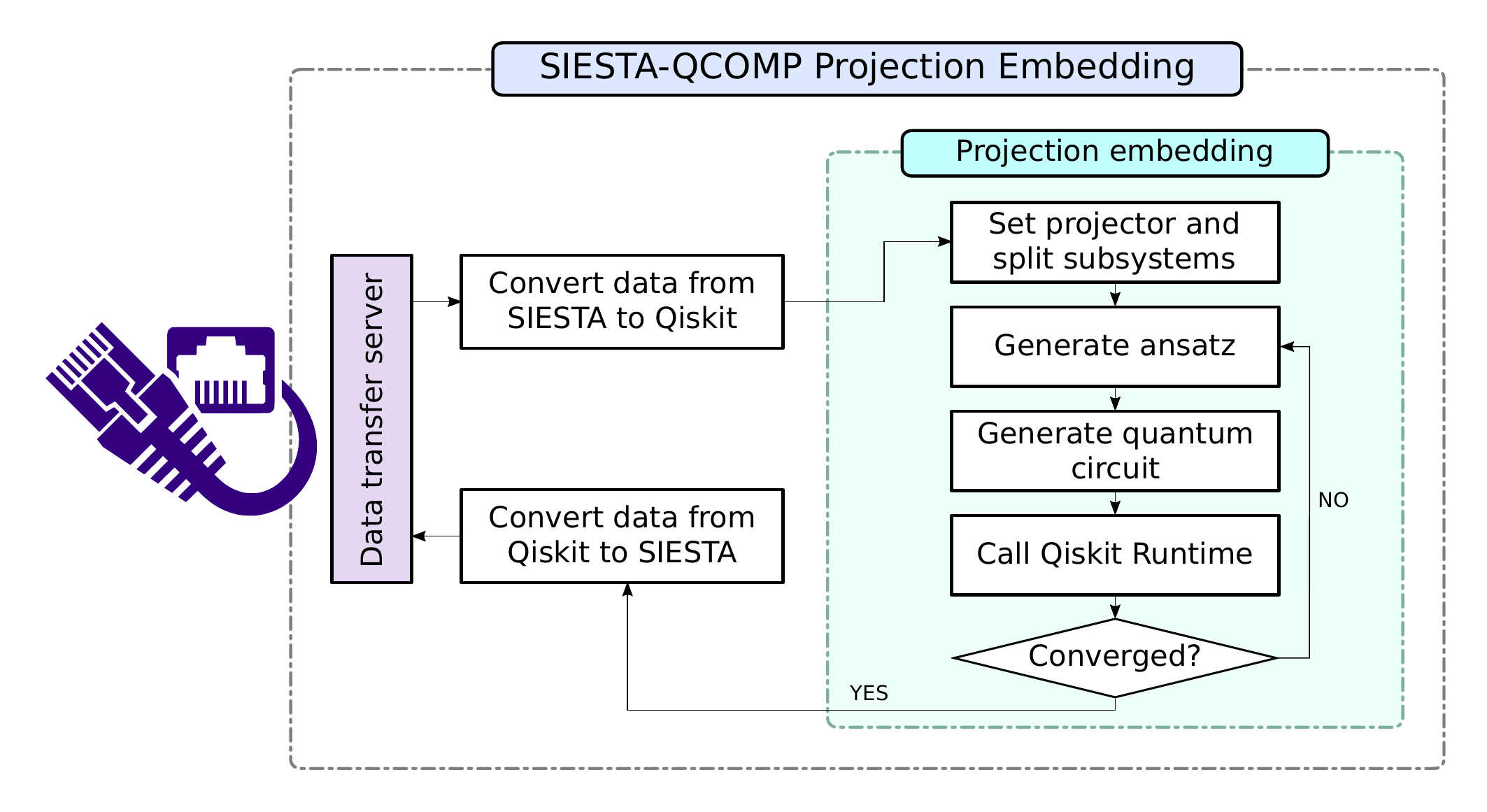}}
  \caption{Embedded part of the hybrid workflow, managed by multiple Python
  components. One component is in charge of the data (left-hand side), while
  the projection embedding and quantum workflow modules (green background)
  organise the execution of the quantum algorithm and the retrieval of its
  results.
  \label{fig:workflow-quantum}}
\end{figure}

At this stage, the problem leaves the realm of materials science coordinates
and enters the realm of fermionic Hamiltonians. The Python driver then runs
the quantum module, which:

\begin{itemize}
  \item Constructs the second-quantised Hamiltonian.
  \item Maps the fermions to qubits (via a Jordan-Wigner or Bravyi-Kitaev
  transformation).
  \item Constructs the ansatz for the quantum algorithm.
  \item Executes the quantum algorithm on the quantum back-end.
  \item Formats the results for their use by the classical engine.
  \item Takes care of the projection-based embedding details.
\end{itemize}

This part of the workflow is described in Fig.~\ref{fig:workflow-quantum}. \\

Upon convergence of the quantum algorithm, the quantum computer returns the
optimised active space density matrix and the ground state energy correction.
These results are sent back through the socket to SIESTA. The latter then
"merges" this quantum correction with the total energy from DFT, producing a
final hybrid energy that accounts for the strong correlations in the active
site (see Fig.~\ref{fig:workflow-classical}). The information contained in
the density matrix of the active space can also be used to improve the
description of the eigenstates by the classical engine. \\

For developers, SIESTA-QCOMP also offers a "file-drop" mode, allowing the
manual execution of each workflow step separately, getting the driver and the
sockets out of the way. While this is not recommended for production, this
execution mode is very useful for testing and debugging, and for tracing the
data flow. It can be used as a fall-back mechanism as well when dealing with
integration issues, e.g. with the batch scheduler of a hybrid
High-Performance Computing (HPC) infrastructure.

\section{Path to Quantum Utility}
\label{sec:quantum-utility}

The ultimate measure of any hybrid quantum-classical architecture is its
ability to address problems that are scientifically significant yet
computationally intractable for standard methods. We are validating the
SIESTA-QCOMP framework by applying it to the study of iron porphyrins, the
active centres of haemoglobin. While VQE served as an excellent workhorse for
demonstrating hybrid workflows on near-term noisy quantum computers, we must
acknowledge its limitations regarding "Quantum Utility" --- the point where a
quantum computer provides a scientific or commercial advantage. Up to now, VQE
has not been very successful beyond 32 qubits, which is considered as the
lower bound towards quantum utility, since it is the upper bound for a
classical supercomputer to simulate a quantum circuit. \\

VQE is a variational method, meaning it relies on a classical optimisation loop
to minimise energy. As system sizes grow, this approach faces two significant
hurdles:

\begin{enumerate}
  \item Measurement Overhead: Estimating the energy requires a vast number of
  measurements (shots) to achieve chemical accuracy, scaling poorly with the
  number of terms in the Hamiltonian.
  \item Optimisation Landscapes: The classical optimisation is plagued with
  barren plateaus --- regions where gradients vanish exponentially, making
  it difficult to find the true ground state of the system.
\end{enumerate}

To move beyond the limitations of standard VQE, the SIESTA-QCOMP road map
prioritises the integration of Sample-based Quantum Diagonalization (SQD)
\cite{Robledo2025}. While VQE relies on a non-linear optimization loop to find
a single ground state, SQD harnesses the quantum processor as a generator of
chemically relevant basis states, which are then processed classically. In
this approach, the quantum computer is used to prepare a trial wavefunction
--- often an approximate ground state obtained from a coarse VQE optimization
or a heuristic ansatz. We then sample this wavefunction in the computational
basis, collecting a set of high-probability electronic configurations
(bitstrings). These sampled configurations define a compact, physically
relevant subspace of the full Hilbert space. The Hamiltonian is then projected
onto this subspace, and the resulting generalized eigenvalue problem is solved
on a classical computer. This hybrid strategy offers two distinct advantages
over pure variational methods:

\begin{itemize}
\item Error Mitigation and Correction: By diagonalising the Hamiltonian in the
subspace of sampled states, SQD can variationally improve upon the initial
quantum state. It can effectively "clean up" certain types of noise or ansatz
deficiencies, recovering the correlation energy that a rigid VQE circuit might
miss.
\item Access to Excited States: Because the final step involves the
diagonalisation of a matrix, SQD naturally yields not just the ground state,
but also a spectrum of excited states within the sampled subspace. This is
particularly valuable for the study of photochemical processes or
spin-crossover transitions in systems like the iron porphyrin complex, where
the energy gap between the singlet and triplet states is the key quantity of
interest.
\end{itemize}

The modular design of our driver (described in Section~\ref{sec:architecture})
makes the transition from VQE to SQD relatively seamless. The interface for
the Fermionic Hamiltonian generation remains almost identical; most changes
affect the post-processing of the quantum measurement data. By shifting the
heavy lifting of diagonalisation back to the classical side --- but performing
it within a "quantum-selected" basis --- SQD has demonstrated it is a robust
path toward Quantum Utility, balancing the sampling power of the QPU with the
numerical precision of classical linear algebra \cite{Robledo2025}. \\

\begin{figure}
  \begin{center}
    \begin{tabular}{c c}
      \includegraphics[width=2.25in]{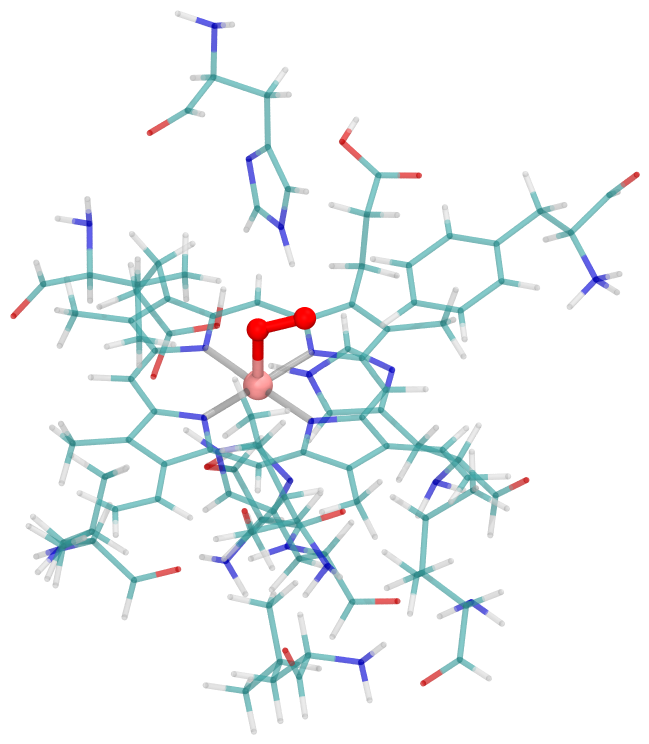} &
      \includegraphics[width=2.25in]{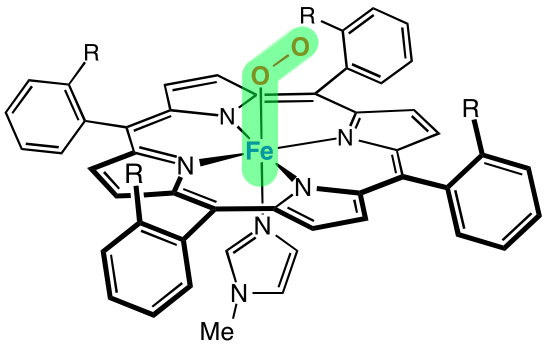} \\
      \mbox{} & \mbox{} \\
      a & b \\
    \end{tabular}
  \end{center}
  \caption{Iron porphyrin complex used to validate the hybrid workflow.
  (a) 3D structure of the iron porphyrin molecule within the haemoglobin
  protein environment, with the iron and oxygen atoms highlighted. Some
  amino-acids have been removed to make the central part of the molecule
  easier to see. (b) Atoms involved in the active space definition, centred
  on the Fe atom and its $3d$ orbitals, and including orbitals from the oxygen
  atoms. This region is treated through quantum computing to resolve the
  competition between different spin states (singlet, triplet, quintet),
  while the surrounding porphyrin ring and protein scaffold are treated
  classically. To simplify the view, we have represented it on a synthetic
  porphyrin.
  \label{fig:porphyrin}}
\end{figure}

Metal-organic chemistry is a field where the limitations of DFT appear very
clearly. In particular, iron porphyrins represent a quintessential
strong-correlation problem in biology. The binding of oxygen to the central
iron atom involves subtle spin-state transitions that dictate the biological
function of the molecule. Standard DFT functionals often struggle to correctly
predict the energy hierarchy of these spin states (e.g., singlet, triplet,
quintet) due to the multi-reference character of the iron $d$-electrons. In
the current stage of our endeavour, the system is partitioned naturally:

\begin{itemize}
  \item The Environment: The complex protein scaffold and the organic porphyrin
  ring are treated efficiently by SIESTA using standard DFT. This captures the
  steric constraints and the electrostatic environment.
  \item The Active Space: The iron centre and its immediate coordination shell
  (the $d$-orbitals) are projected out and mapped to the quantum processor.
  When bound to a gas molecule like O$_2$, CO, or CO$_2$, we can also include
  the electrons and orbitals of this molecule into the active space, to get
  insight on their binding to the iron atom.
\end{itemize}

By solving this impurity problem with SQD, we aim to recover the static
correlation energy missing from the DFT description. This test case serves not
only as a validation of the general SIESTA-QCOMP workflow but also as a
blueprint for future studies of compounds where the electronic structure of a
localised area dictates the catalytic activity of a larger system. At the time
this article is written, we have performed multiple series of DFT calculations
for different porphyrin sizes and the gas molecules listed above. This leads
to a "test matrix" of 16 individual combinations of the heme center and gas
molecules (as the one in Fig.~\ref{fig:porphyrin}b) and between 0 and 3 layers
of amino-acids around it, as shown in Fig.~\ref{fig:siesta-qcf}. Extracting
the amino-acid layers from hemoglobin, fixing the dangling bonds and redox
states, and adding the missing hydrogen atoms, was done with the VMD
\cite{Hump1996} and SAMSON \cite{SAMSON2025} software packages. \\

\begin{figure}
  \begin{center}
    \begin{tabular}{|l|c|c|c|c|}
      \hline
      \textbf{Porphyrin} & \textbf{Bare} & \textbf{O$\mathbf{_2}$} &
      \textbf{CO} & \textbf{CO$\mathbf{_2}$} \\
      \hline
      Heme alone & 00 & 01 & 02 & 03 \\
      \hline
      Heme + 1 layer & 10 & 11 & 12 & 13 \\
      \hline
      Heme + 2 layers & 20 & 21 & 22 & 23 \\
      \hline
      Heme + 3 layers & 30 & 31 & 32 & 33 \\
      \hline
    \end{tabular}
  \end{center}
  \caption{SIESTA-QCOMP test matrix for the iron porphyrin of hemoglobin. Each
  test is identified by a 2-digit number: the first digit is the number of
  amino-acid layers around the central pyrrole ring, and the second digit
  indicates which gas molecule is attached to it. A zero means no amino-acid
  or no gas molecule.
  \label{fig:siesta-qcf}}
\end{figure}

We are now in the process of testing the whole SIESTA-QCOMP workflow with
systems 00, 01, 10, and 11, before addressing the other ones. Our main
objective with the whole series of tests is to identify how the main
parameters of the active space and its environment influence the results of
the hybrid calculation. We will then be able to propose a streamlined workflow
for this kind of system to the users of SIESTA-QCOMP. The whole software
suite will be made available progressively on Gitlab, at
\url{https://gitlab.com/qcomp4dft/}, under the Apache License Version 2.0.

\section{Discussion: The QCOMP4DFT Vision}
\label{sec:qcomp4dft}

The development of SIESTA-QCOMP is not an isolated effort but a foundational
step toward a broader initiative: QCOMP4DFT (Quantum Computing for Density
Functional Theory). As we reflected during the Royal Society discussion
meeting, the challenge of integrating quantum computing into materials science
is too vast for any single group or code to solve alone. Our approach offers
distinct advantages:

\begin{itemize}
  \item Language Agnosticism: The quantum back-end is independent of the
  DFT code. While our current implementation uses Qiskit, the driver can
  easily be adapted to other frameworks without modifying SIESTA.
  \item Interoperability: The socket protocol establishes a standard for data
  exchange (Hamiltonian parameters out, density matrix in) that can
  potentially be adopted by other electronic structure codes, paving the way
  for the QCOMP4DFT vision.
\end{itemize}

Historically, electronic structure codes have been "monolithic fortresses" ---
massive, self-contained packages optimised for high-performance computing
(HPC). Noticeable and impactful efforts have been conducted towards higher
modularity \cite{Oliveira2020}, but progress has been slow. The quantum era
demands a rapid paradigm shift towards modular ecosystems. The diversity of
quantum hardware (superconducting qubits, trapped ions, neutral atoms,
photonic qubits) and the rapid turnover of quantum software frameworks mean
that rigid coupling between a DFT code and a specific quantum back-end is
destined for obsolescence. \\

By labelling our initiative as "QCOMP4DFT", we express our intention to use
QPUs as enhancers to DFT calculations, similarly to what
GPUs are doing with CPUs. This is
different from other works exploring the use of quantum computers to build
DFT Hamiltonians \cite{Baker2020,Senjean2023}. In the electronic structure
community, there is a whole ecosystem currently emerging around the use of
QPUs, with a growing number of codes and libraries cross-interfacing with each
other. To give just a few examples: Qiskit interfaces with Psi4
\cite{Smith2020} ,
PySCF \cite{Sun2020}, and PyQuante \cite{Muller2004}; OpenFermion
\cite{McClean2020} interfaces with Dirac
\cite{Saue2020}, Psi4, PySCF, and Q-CHEM \cite{Shao2015}; Tequila
\cite{Kottmann2021} interfaces with
Madness \cite{Harrison2016}, Orca \cite{Neese2012}, Psi4, and PySCF. Among all these
multiple integration efforts, recurrent architectural patterns appear. \\

\begin{figure}
  \centerline{\includegraphics[width=4.5in]{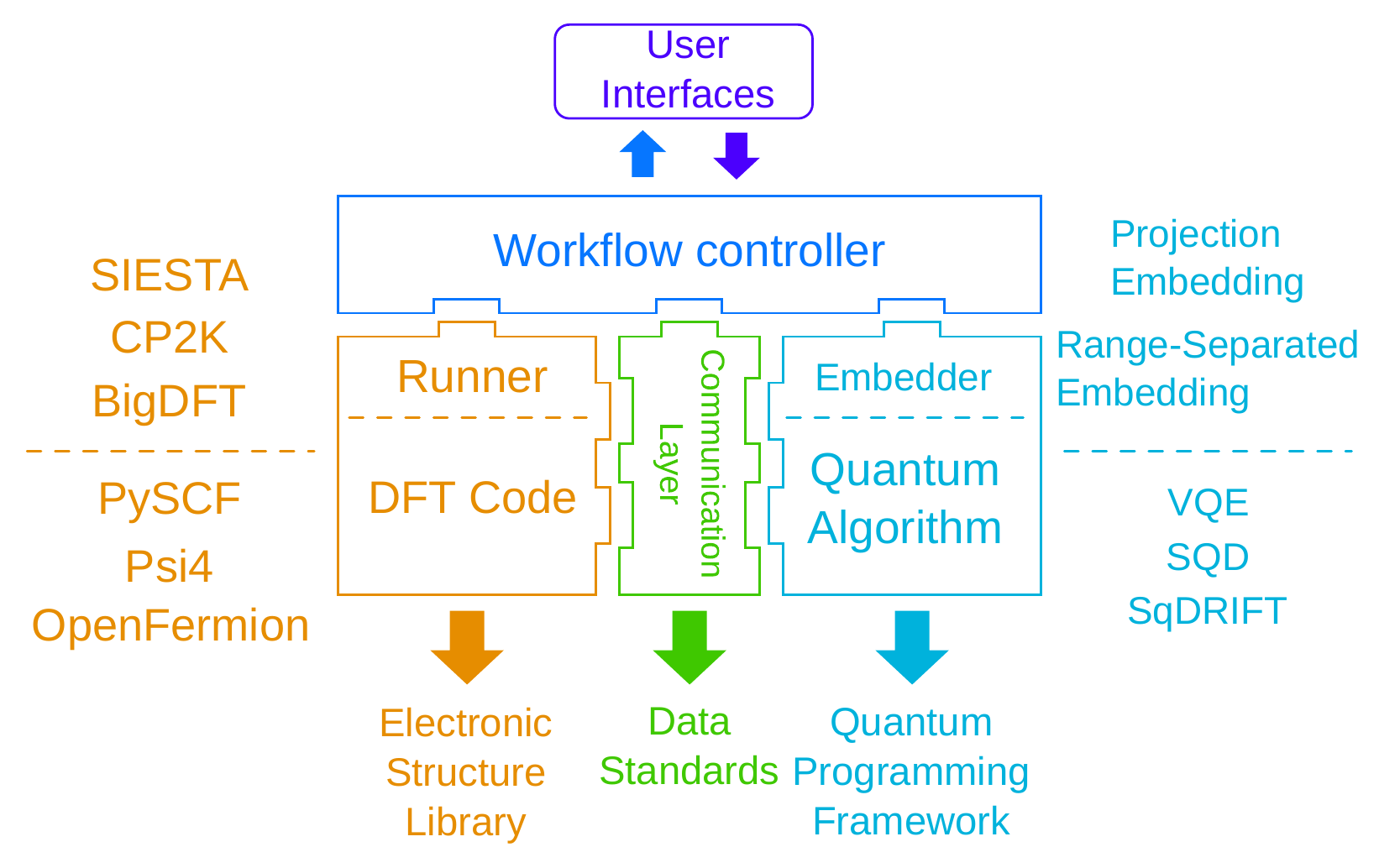}}
  \caption{Proposed software architecture for hybrid quantum-classical
    workflows. On the left, the classical module (orange) is structured as
    a wrapped DFT code, with a Python runner to fine-tune the input and manage
    the output. On the right, the quantum module (light blue) is implemented
    as a quantum algorithm bound to an embedder. Examples of software packages
    for the classical module are provided on the left, with our confirmed
    targets for production calculations (top) and the packages we use for test
    calculations (bottom). Examples of embedding techniques (top) and quantum
    algorithms (bottom) are provided on the right. The main difference with
    standard HPC calculations is the presence of a workflow controller
    orchestrating the different steps. The three modules below the controller
    are a generalisation of the architecture presented in this article.
    \label{fig:qcomp4dft-arch}}
\end{figure}

The socket-based interface designed in SIESTA-QCOMP serves as a prototype
for this new paradigm. It suggests a future where the "Quantum Embedding"
layer acts as a standardised middleware. In this vision, illustrated in
Fig.~\ref{fig:qcomp4dft-arch}, a user could seamlessly switch the classical
engine (from SIESTA to CP2K to BigDFT) or the quantum back-end (from Qiskit to
another platform) without rewriting the workflow, provided they adhere to a
common data exchange protocol for Hamiltonians and density matrices. \\

Perhaps the most critical bottleneck identified during the workshop is human,
not technical. The "languages" spoken by quantum engineers (qubits, gates,
error mitigation and correction) and researchers (functionals, basis sets,
orbitals) are often mutually unintelligible. Success in this field requires
deep collaboration. SIESTA-QCOMP was born from such a dialogue --- merging the
specific domain knowledge of electronic structure theory with the algorithmic
expertise of quantum information science. The QCOMP4DFT initiative aims to
formalise this collaboration, building a community where standard interfaces
allow domain experts to utilise quantum acceleration without needing to become
experts in quantum circuit design and transpilation.

\section{Conclusion}
\label{sec:conclusion}

We stand at a unique juncture in the history of computational science. The
promise of quantum computing to solve the "strong correlation problem" is
theoretically sound, yet practically constrained by the noise and size of
current devices. In this era of near-term noisy quantum computers, hybrid
architectures are not merely a stopgap; they are the necessary bridge to the
future. In this paper, we have presented SIESTA-QCOMP, a realisation of this
hybrid philosophy. By embedding high-level quantum solvers into the robust,
localised-basis framework of SIESTA via a modular socket interface, we are
creating a flexible tool capable of targeting the most challenging electronic
structures in chemistry and physics. From the specific implementation of
projection embedding to the simulation of iron porphyrins, and looking forward
to dynamic solvers like SQD and SqDRIFT, this work intends to establish a road
map for the field. It supports the vision that, by coupling the "best of both
worlds" --- the scalability of DFT and the accuracy of quantum chemistry
methods, we can begin to explore a new frontier in materials design, paving
the way for the era of fault-tolerant quantum utility, as well as multi-tiered
hybrid approaches such as "Quantum Computing (or Quantum Chemistry) in DFT in
Molecular Mechanics" (QC/DFT/MM).

\ack{EA, FM and YP acknowledge funding by MCIN/AEI/10.13039/501100011033
through grant PID2022-139776NB-C65, as well as a Mar\'{\i}a de Maeztu award to
Nanogune, Grant CEX2020-001038-M, and the European Union
NextGenerationEU/PRTR-C17.I1 grant and by IKUR Strategy under the
collaboration agreement between Donostia International Physics Center and CIC
nanoGUNE, on behalf of the Department of Education of the Basque Government
(IKUR QT 2023-2024 and IKUR-QT-SIESQISQD); by the Basque Government and
Ikerbasque Foundation through BasQ project SIESTA-QCF; by Agencia Estatal de
Investigaci\'{o}n, Ministerio de Ciencia e Innovaci\'{o}n, Spain (Proyectos de
Generaci\'{o}n de Conocimiento 2022 program, PID2022-139776NB-C65).
EA acknowledges the United Kingdom's EPSRC (Grant EP/V062654/1). FM and YP
thankfully acknowledge the computer resources at and technical support from
DIPC (Hyperion), and MareNostrum, and the technical support provided by
Barcelona Supercomputing Center (FI-2025-2-0068); Simune Atomistics for access
to a development version of their Atomistic Simulation Advanced Platform
(ASAP). YP gratefully thanks M.~Rossmannek for a very useful and practical
introduction to hybrid quantum-classical embedding, very interesting
discussions, and guidance to run quantum algorithms; S.~Pf\"{a}ndler for
comprehensive guidance and information regarding access to and use of quantum
computers; L.~Reining, A.~Carreras, D.~Casanova, and the Basque Quantum
Community, for many insightful and stimulating discussions; M.~Sauban\`{e}re
for weaving collaborations towards new DFT codes and new quantum algorithms.}

%%%%%%%%%% Insert bibliography here %%%%%%%%%%%%%%

\bibliographystyle{IEEEtran}
\bibliography{siesta-qcomp}
\nocite{*}

\end{document}